\documentclass[journal]{IEEEtran}
\usepackage{amsmath}
\usepackage{algorithm}
\usepackage{algorithmic}
\usepackage{subfigure}
\usepackage{cite}
\usepackage{makecell}
\usepackage{url}
\usepackage{slashbox}
\usepackage[breaklinks]{hyperref}
\usepackage{epsfig,psfig}
\usepackage{graphicx}
\usepackage{dblfloatfix}
\usepackage{caption}

\begin{document}

\title{Hierarchical Network Data Analytics Framework for B5G Network Automation: Design and Implementation}

\author{Youbin Jeon and Sangheon Pack,~\IEEEmembership{Senior Member,~IEEE} }
\maketitle

\begin{abstract}
5G introduced modularized network functions (NFs) to support emerging services in a more flexible and elastic manner. To mitigate the complexity in such modularized NF management, automated network operation and management are indispensable, and thus the 3rd generation partnership project (3GPP) has introduced a network data analytics function (NWDAF). However, a conventional NWDAF needs to conduct both inference and training tasks, and thus it is difficult to provide the analytics results to NFs in a timely manner for an increased number of analytics requests. In this article, we propose a hierarchical network data analytics framework (H-NDAF) where inference tasks are distributed to multiple leaf NWDAFs and training tasks are conducted at the root NWDAF. Extensive simulation results using open-source software (i.e., free5GC) demonstrate that H-NDAF can provide sufficiently accurate analytics and faster analytics provision time compared to the conventional NWDAF.
\end{abstract}


\IEEEpeerreviewmaketitle

\section{Introduction}

To meet emerging service requirements in various industries (e.g., autonomous driving and smart factory), the 3rd generation partnership project (3GPP) introduced a new fifth-generation (5G) system architecture. Unlike the fourth-generation (4G) long-term evolution (LTE) system, 5G defines modularized network functions (NFs) for flexible network management and operation, which is beneficial for supporting diverse requirements of new 5G services~\cite{TS23.502}. However, as the number of connected devices rapidly increases and quite diverse applications are considered in 5G, manual network management and operation become intractable. Accordingly, network automation using machine learning (ML) and artificial intelligence (AI) technologies is receiving great attention 6G~\cite{Sevgican20,Liu23}.

In this context, 3GPP has defined a novel NF called network data analytics function (NWDAF)~\cite{TS23.501}, which collects data from other NFs or operation, administration, and management (OAM) systems, and derives the analytics. The analytics can be returned to other NFs for their autonomous operation. For example, NWDAF can obtain some data such as a list of NF instances and their resource usages from NF repository function (NRF)~\cite{TS23.501}. By analyzing these data, NWDAF can construct the NF load analytics and provide it to NF instances~\cite{Jeon22}. Then, these analytics can be used to automatically scale up or down the capacity of NF instances.

NWDAF handles inference tasks as well as training tasks. Unlike training tasks that can be run offline, inference tasks need to be performed online and some of them have strict latency requirements of tens to hundreds of milliseconds per query~\cite{zhang19,Nauman22}. However, in a conventional NWDAF, both training and inference tasks are processed by a monolithic entity, which consumes most of resources only for training tasks. Therefore, inference tasks can be further delayed. A recent proposal of 3GPP~\cite{TS23.288} introduced decomposition of the NWDAF function into training and inference ones. However, it is difficult for a single inference function to provide timely analytics results when a number of NFs request different types of analytics (e.g., UE mobility prediction and NF load analytics). To address this problem, an improved NWDAF architecture that provides sufficiently accurate analytics, as well as faster analytics time, is strongly required in 6G.

In this article, we propose a hierarchical network data analytics framework (H-NDAF) that consists of root and leaf NWDAFs for 6G. The root NWDAF is a fully functional NWDAF that can conduct data collection, model training, and inference. On the other hand, the leaf NWDAF collocated at each NF is a lightweight NWDAF having only the inference module. In H-NDAF, the root NWDAF can construct various models with improved accuracy based on the global data and these models are delivered to the leaf NWDAFs. Then, analytics requests from a specific NF can be locally handled by its collocated the leaf NWDAF, which guarantees faster analytics provision time (i.e., inference results can be provided in a timely manner). To show the feasibility of H-NDAF, we implemented H-NDAF using free5GC~\cite{free5GC}, which is a well-known open-source software for a 5G core network (CN), and released at \url{https://github.com/youbinee/H-NDAF}. Evaluation results on the UE throughput prediction demonstrate that H-NDAF provides sufficiently high accuracy and fast analytics provision time.

\begin{figure*}%[!t]
\centering
\includegraphics[width=6in]{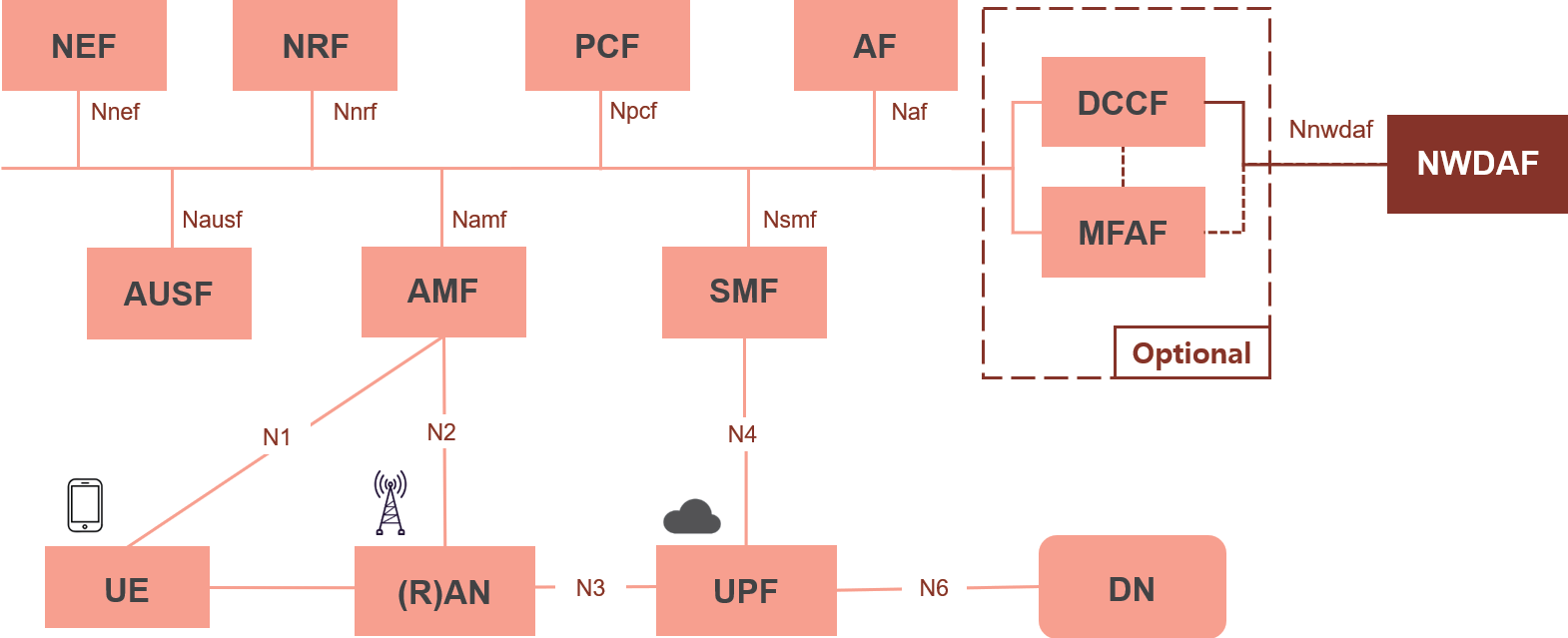}
\caption{5G network architecture.}
\label{fig:fig_1}
\end{figure*}

The rest of this article is organized as follows. The backgrounds on the 5G network architecture and NWDAF are first introduced. After that, H-NDAF is described with its detailed components and a representative use case. Then, the implementation and simulation results are presented and followed by open research issues in H-NDAF. Finally, the concluding remarks are given.

\section{Background}
\label{Sec:backandmotiv}
In this section, we describe the 5G network architecture and detail NWDAF for network automation.

\subsection{5G Network Architecture}

The 5G network consists of user equipment (UE), radio access network (RAN), and CN. Recently, UE goes beyond traditional smartphones and tablets to include a wider range of vehicles, drones, industrial machines, robots, home appliances, medical devices, and so on. RAN is responsible for transmission between UEs and base stations (i.e., gNBs) and handles efficient use of the radio spectrum. CN provides a bridge between RAN and Internet, and consists of different types of NFs. NFs can be classified into 1) control plane (CP) NFs in charge of network management and operation (e.g., UE authentication, mobility management, and session management); and 2) user plane (UP) NF in charge of data transmissions (e.g., routing and forwarding).

As shown in Figure~\ref{fig:fig_1}, 5G CP is designed as a novel service-based architecture (SBA) where NFs have their own functionalities (i.e., services) and expose them through a service-based interface (SBI) (i.e., RESTful interface using HTTP/2)~\cite{Alawe18}. Each NF defines its own interface to a common service bus (i.e., per-NF interface) for exposing its services. For example, \textsf{Namf} represents the interface to the access and mobility management function (AMF) to the common service bus, which is used to send control messages to/from AMF. The functionalities of key NFs in CP are as follows. AMF manages registration and mobility management, and the session management function (SMF) controls session requests and IP address management. In addition, the policy control function (PCF) is to manage and provide policy rules to other NFs, and the authentication server function (AUSF) is responsible for the security procedure for UE authentication. Also, the application function (AF) exposes application services to UE and interacts with PCF for its policy control. The network exposure function (NEF) and NRF are for vertical service integration and NF management, respectively. In detail, NEF securely opens the network functionalities to third-party applications. On the other hand, NRF enables service discovery and status updates of NFs in CN. For this, all NFs in CN register their provided services to NRF through the service registration procedure. Then, they can use NRF as a database to search for services provided by other NFs through service discovery and service approval procedures.

In 5G UP, the user plane function (UPF) delivers data traffic from/to Internet and is responsible for reporting traffic usage and applying an appropriate quality of service (QoS) policy. The interaction between CP and UP is conducted based on the traditional point-to-point (P2P) interfaces. For example, the interfaces \textsf{N1}, \textsf{N2}, and \textsf{N3} are defined for UE-AMF, gNB-AMF, and gNB-UPF interactions, respectively.

\begin{figure*}%[!t]
\centering
\includegraphics[width=6in]{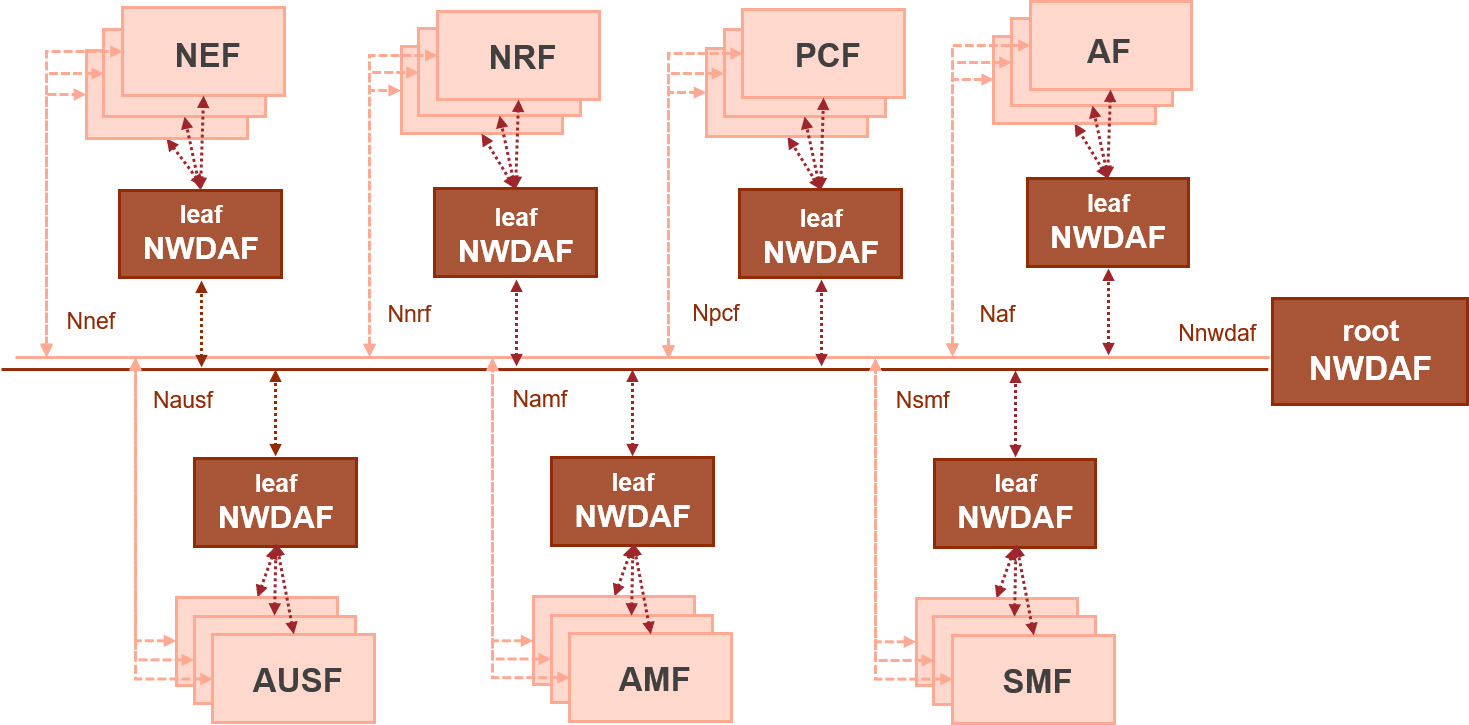}
\caption{H-NDAF architecture.}
\label{fig:fig_2}
\end{figure*}

\subsection{NWDAF}

NWDAF aims to support intelligent and autonomous network operation and service management. For this, NWDAF can provide analytics to other NFs based on two types of service provision methods: 1) analytics request method; and 2) analytics subscription method. In the analytics request method, a specific NF requests the analytics to NWDAF through the \textsf{Nnwdaf} interface, and then NWDAF delivers the analytics to that NF. In the analytics subscription method, NF subscribes its interest in specific analytics to NWDAF. Then, whenever the corresponding analytics results are ready or updated, they are delivered to NF. For both service methods, NWDAF performs the following procedures.

First of all, NWDAF collects a variety of data from NFs as well as OAM systems. For efficient data collection and management, two optional NFs, i.e., data collection coordination function (DCCF) and messaging framework adapter function (MFAF), can be deployed~\cite{TS23.288}. DCCF coordinates the data collection between multiple data sources (e.g., NFs) and NWDAFs to reduce the traffic volume for the data collection (or to avoid redundant data collections). Specifically, DCCF can decide 1) whether to deliver data to NWDAF or not and 2) whether multiple data are combined and delivered as a single message or not. In addition, it can extract and summarize multiple data to create a consolidated data summary. Meanwhile, MFAF can be exploited to format or process data before delivering data to NWDAF.

After sufficient data has been collected, the model training can be performed. Specifically, a model training logical function (MTLF) of NWDAF can train ML/AI models (e.g., neural networks). Note that, since the data is collected continuously, MTLF can periodically update models. After constructing models, whenever NFs request the analytics, an analytics logical function (AnLF) of NWDAF can generate the analytics results (i.e., conduct the inference using models) and deliver them to the requesting NFs. Based on the analytics, NFs can optimize and improve their operations. Meanwhile, if NFs subscribe to the analytics, AnLF can periodically provide the analytics results to them.

\begin{figure*}%[!t]
\centering
\includegraphics[width=6in]{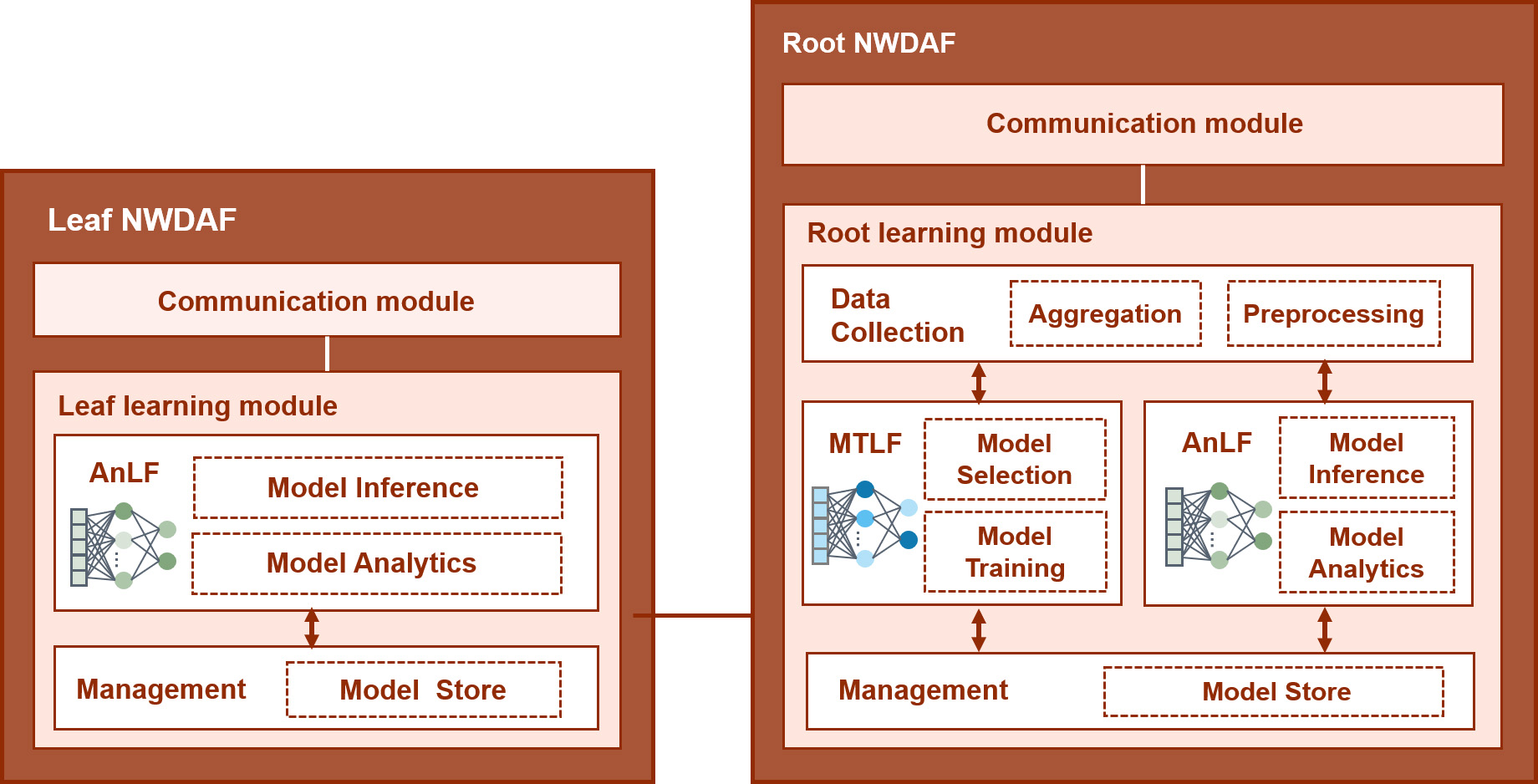}
\caption{Internal components of root/leaf NWDAFs.}
\label{fig:fig_3}
\end{figure*}

\section{H-NDAF}
\label{Sec:hierarchicalNWDAF}

To provide sufficiently accurate analytics services in a timely manner, we design H-NDAF. In this section, we describe the details of H-NDAF including a specific use case.

\subsection{Architecture and Operating Procedure}
 
Figure~\ref{fig:fig_2} shows the overall architecture of H-NDAF, in which there are single root NWDAF and multiple leaf NWDAFs collocated at NFs. As shown in Figure~\ref{fig:fig_3}, the root NWDAF consists of communication and learning modules. The communication module is responsible for communicating with other NFs (e.g., NRF registration) and leaf NWDAFs (e.g., model delivery). The learning module can be divided into four sub-modules: data collection, MTLF, AnLF, and model store. The data collection sub-module can collect data from multiple NFs, AFs, and OAM in a network-wide view (i.e., global view). In addition, it can conduct data pre-processing (e.g., normalization) for efficient training. Next, the MTLF sub-module selects appropriate models (e.g., convolutional neural network (CNN) and recurrent neural network (RNN)) by considering the data types and/or the purpose of the models, and then trains the models by using the collected data. The AnLF sub-module utilizes models generated by MTLF to derive analytical results through inference. Meanwhile, as shown in Figure~\ref{fig:fig_3}, the leaf NWDAF consists of communication and learning modules. The communication module is responsible for the connection to the collocated NF and root NWDAF. Unlike the root NWDAF, the learning module in the leaf NWDAF includes only AnLF and model store sub-modules. Therefore, to derive the analytics results in the AnLF sub-module, the leaf NWDAF should receive the model from the root NWDAF.

In H-NDAF, the analytics services can be provided to NF by either 1) the request method or 2) the subscription method as in the standard~\cite{TS23.288}. In the analytics request method, when NF sends an analytics request message to the leaf NWDAF, the leaf NWDAF checks whether a model providing the requested analytics exists in its storage. If the leaf NWDAF has the model, the inference is conducted immediately. Meanwhile, if the leaf NWDAF does not have any appropriate model, the leaf NWDAF requests the model to the root NWDAF. On receiving the request, the root NWDAF searches the model in its storage and transmits it to the leaf NWDAF. After that, the leaf NWDAF can use the model for inference tasks, and the analytics (i.e., inference results) is delivered to the requesting NF.

In the analytics subscription method, NF sends an analytics subscription message to the leaf NWDAF. Then, the leaf NWDAF sends the model subscription message to the root NWDAF. After receiving the model subscription message, the root NWDAF periodically transmits the model to the leaf NWDAF. Then, the leaf NWDAF stores the model and periodically delivers the analytics results to the requesting NF. Later, NF can send an analytics unsubscription message to the leaf NWDAF. Then, the leaf NWDAF deletes the model and sends the analytics unsubscription message to the root NWDAF.

In H-NDAF, a leaf NWDAF has a limited capacity to store requested and subscribed models. Therefore, H-NDAF introduces criteria to store the models depending on the types of the methods (i.e., request and subscription methods). In the subscription method, the analytics services are regularly provided to NFs. Meanwhile, in the request method, different analytics services are irregularly requested and some of them may not be exploited for a long time. Therefore, higher priority is given to the subscribed models than the requested models in storing them at the model storage. In addition, models with higher frequency (i.e., higher popularity in using the models) among the subscribed models (or requested models) have higher priority than those with lower frequency. When the model storage is fully occupied, H-NDAF applies the following replacement policy. If a new model is a subscribed one, it is stored after removing an existing one with the lowest frequency among the saved requested models. On the other hand, if a new model is a requested one, it can be stored only when its frequency is higher than the lowest frequency of the existing requested models.

\begin{figure*}%[!t]
\centering
\includegraphics[width=6in]{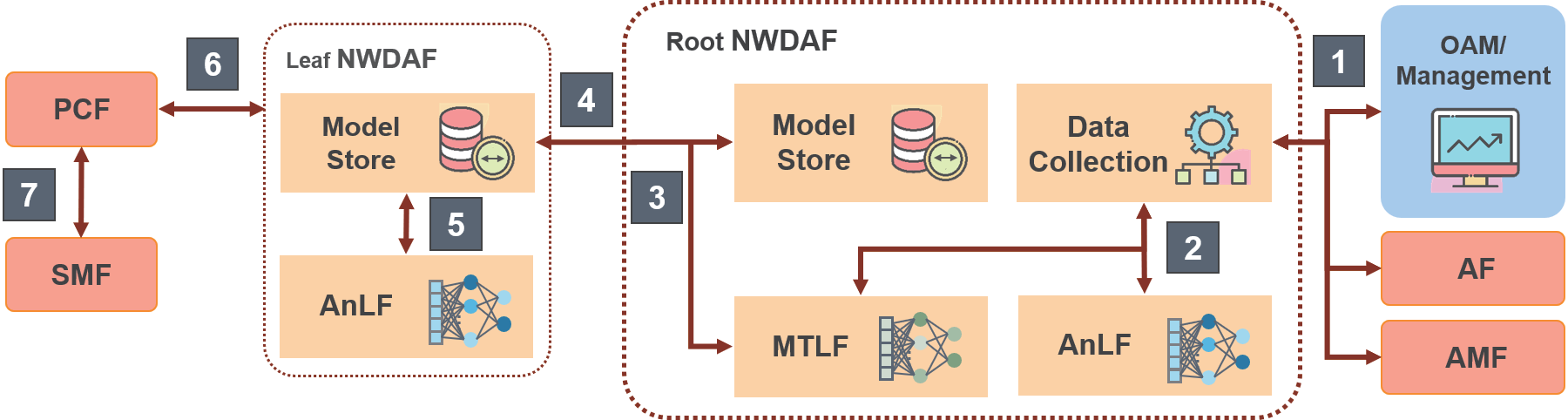}
\caption{Use case of UE throughput prediction in H-NDAF.}
\label{fig:fig_4}
\end{figure*}

\subsection{Use Case}
\label{Sec:usecases}

The analytics services listed in 3GPP standard~\cite{TS23.288} can be classified into three types: 1) UE-related analytics; 2) CN-related analytics; and 3) end-to-end related analytics. UE-related analytics includes UE mobility analytics (e.g., UE registration area management and paging area management) and UE communication analytics (e.g., traffic routing handling). By means of these analytics, it is possible to improve QoS in UE communications. CN-related analytics includes NF load analytics and slice load analytics. Through these analytics, the load on CN can be flexibly managed by adjusting the capacity of NF instances and/or slices. Lastly, end-to-end related analytics includes QoS sustainability analytics for QoS change in a specific region and data network (DN) performance analytics in terms of throughput and latency.

We describe a use case of how PCF in H-NDAF exploits the analytics on the UE downlink throughput prediction (i.e., the analytics on how the achievable data rate of UE changes) to optimize the policy for UE data flows. In this use case, the root NWDAF first collects raw data such as UE location from AMF, RAN status from OAM, and UE ID from AF (step $1$ in Figure~\ref{fig:fig_4}). After that, the root NWDAF aggregates the collected raw data and conducts the preprocessing to make them in a trainable form. Then, the preprocessed data is delivered to MTLF to train and generate a UE throughput prediction model (step $2$ in Figure~\ref{fig:fig_4}). The model is saved at the model store of the root NWDAF (step $3$ in Figure~\ref{fig:fig_4}). When the leaf NWDAF of PCF requests the model (or subscribes the model) for the UE throughput prediction, the root NWDAF transmits the trained model in its model store to the leaf NWDAF (step $4$ in Figure~\ref{fig:fig_4}). Then, the leaf NWDAF can store the trained model in its model store and derive the analytics by using its AnLF sub-module (step $5$ in Figure~\ref{fig:fig_4}). The derived analytics results on the UE throughput prediction are transmitted to PCF collocated with the leaf NWDAF (step $6$ in Figure~\ref{fig:fig_4}). Depending on the received analytics results, PCF updates the policy for UE data flows (step $7$ in Figure~\ref{fig:fig_4}). For example, if the UE throughput is expected to decrease based on the analytics, PCF can assign a higher QoS level to UE. In addition, to guarantee the updated QoS level, UPF can reserve more resources for UE, which can lead to a smooth data flow to it.

In H-NDAF, it is possible to accommodate a variety of models with a limited model store by reusing the trained model for multiple services. For example, a model for the UE throughput prediction can be reused for load balancing and abnormal behavior detection services. Specifically, given the throughput prediction models for multiple UEs, a situation, where an excessive number of UEs are concentrated in a single UPF, can be detected. In this situation, load balancing can be triggered by distributing these UEs to multiple UPFs before severe throughput degradation. In addition, the UE throughput prediction model can catch another situation where the UE throughput is expected to abruptly decrease due to denial-of-service (DoS) attacks. When detecting this abnormality, SMF can adjust the maximum bit rate (MBR) of the attacker to prevent the throughput reduction of other legal UEs.

\section{Performance Evaluation}
\label{Sec:performance}

For performance evaluation, we have implemented a prototype of H-NDAF based on free5GC~\cite{free5GC}, which is a well-known 5G open-source project being developed by National Chiao Tung University in Taiwan. The current public version of free5GC includes several core NFs such as AMF and SMF. However, it does not include NWDAF, and thus we implemented the leaf and root NWDAFs in free5GC for H-NDAF.

To evaluate the performance of H-NDAF, we consider the following two comparison frameworks: 1) CONV where there is a single NWDAF for whole NFs; and 2) MULTI where multiple NWDAFs are deployed and each one covers three NFs. We consider the analytics on the UE throughput prediction described in the previous section. To make the UE throughput prediction model, we used the Lumos5G~\cite{lumos5g} dataset including 5G throughput measurements in real environments. The dataset consists of UE ID, RAN status, UE location, signal strength, past throughput, and current throughput. The first five elements (i.e., UE ID, RAN status, UE location, signal strength, and past throughput) are exploited as an input vector of the model, and the last element (i.e., current throughput) is used as an output of the model. Meanwhile, the long short-term memory (LSTM)~\cite{lstm} model is adopted since it is suitable for time-domain prediction to learn long-term relationships. Specifically, we use a three-layer LSTM architecture with $128$ hidden units and mean-squared-error (MSE) as a loss function. H-NDAF runs on a single machine with Intel Core i7-10700 CPU and NVIDIA GEFORCE RTX 2070. The source code of H-NDAF can be found at \url{https://github.com/youbinee/H-NDAF}, which gives detailed instructions to run H-NDAF.

\begin{figure*} 
\centering
    \subfigure[]{
    \includegraphics[width=0.98\columnwidth]{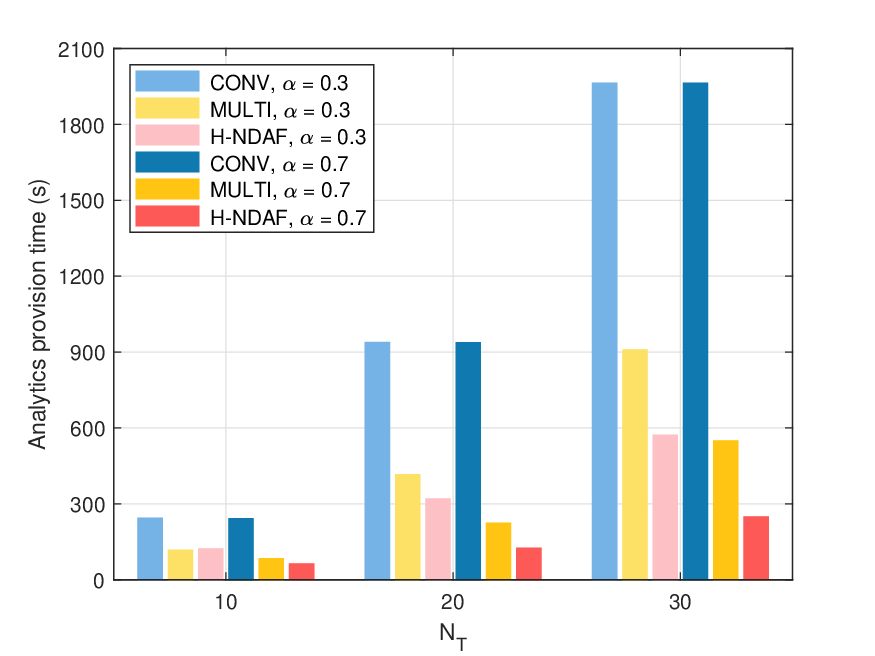}
    }
    \subfigure[]{
    \includegraphics[width=0.98\columnwidth]{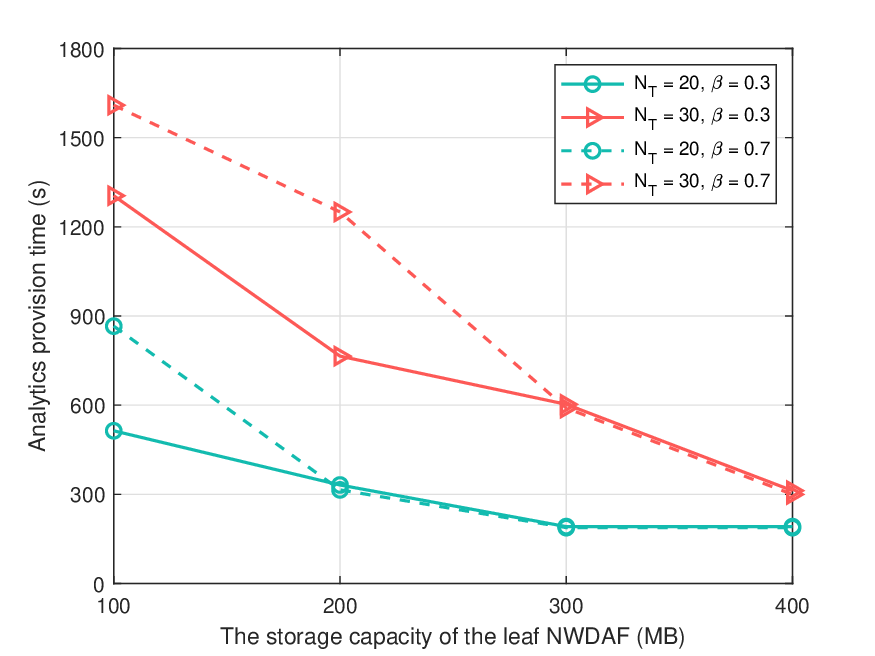}
    }
    \caption{Analytics provision time: (a) effect of $N_T$ and $\alpha$; (b) effect of storage capacity and $\beta$.}
\label{fig:fig_5}
\end{figure*}

To investigate the effectiveness of the model reuse concept, we define parameters $\alpha$ and $\beta$. $\alpha$ is the probability of requesting a previously used model (i.e., model reuse probability). For example, if $\alpha$ is $0.3$, the previous models are reused with the probability of $0.3$ whereas request messages for new analytics are generated with the probability of $0.7$. Meanwhile, $\beta$ is the ratio of the request method to the total ones. For instance, if $\beta$ is $0.3$, the analytics services are provided by the request method and the subscription method with the probabilities of $0.3$ and $0.7$, respectively.

\subsection{Prediction Accuracy}

We first evaluate the accuracy of the UE throughput prediction model according to the dimension $D$ of the input vector of the LSTM model. For $D=5$, the input vector consists of the UE ID, RAN status, UE location, signal strength, and past throughput. Meanwhile, for $D=3$, the UE location and RAN status are excluded. In terms of the prediction accuracy, the values of MSE, mean absolute errors (MAE), and root mean squared errors (RMSE) are $0.29$, $0.33$, and $0.53$, respectively, when $D = 3$. On the other hand, when $D=5$, they are reduced to $0.14$, $0.21$, and $0.34$, respectively. These results imply that sufficient types of input data should be exploited to make models with higher accuracy. Note that, compared to MULTI, H-NDAF and CONV can create a model with a global view (i.e., all types of input data can be collected by the root NWDAF), and therefore they have  benefits of constructing a model with higher accuarcy.

\subsection{Analytics Provision Time}

Figure~\ref{fig:fig_5}(a) shows the analytics provision time according to the total number of requests/subscriptions (denoted by $N_T$) and the model reuse probability $\alpha$. Note that the analytics provision time is defined as the elapsed time to receive the results of all analytics requests. In this evaluation, it is assumed that all needed models have been already trained, and the model size is $15$MB. Also, NF requests analytics services every $5$ seconds and $\beta=0.5$ (i.e., the portion of request and subscription methods are identical). From Figure~\ref{fig:fig_5}(a), it can be found that, as $N_T$ increases from $10$ to $30$, the analytics provision times of CONV and MULTI increase significantly by $7.1$ and $6.7$ times, respectively. Meanwhile, it can be seen that the analytics provision time of H-NDAF increases only by $3.6$ times under the same situation. In H-NDAF, the models for the requested analytics can be stored and reused in the leaf NWDAFs attached to NFs. In this situation, there is no need to retrieve the models from the root NWDAF and the analytics provision time can be reduced.

From Figure~\ref{fig:fig_5}(a), it can be also seen that the analytics provision time of MULTI and H-NDAF decreases with the increase of $\alpha$. This is because the probability that the model for the requested analytics has been already stored at the distributed leaf NWDAFs increases according to $\alpha$. On the other hand, the analytics provision time of CONV is constant regardless of $\alpha$ since it processes all requests and subscription messages in a single NWDAF.

Figure~\ref{fig:fig_5}(b) shows the analytics provision time of H-NDAF depending on the storage capacity of the leaf NWDAF and the ratio of the request method, $\beta$. The model reuse probability $\alpha$ for $N_T$ is set to $0.5$. As shown in Figure~\ref{fig:fig_5}(b), the analytics provision time of H-NDAF decreases as the storage capacity of the leaf NWDAF increases. This is intuitive because a larger storage capacity indicates that more models can be stored and reused, which implies that the leaf NWDAF does not need to request the models to the root NWDAF. In addition, from Figure~\ref{fig:fig_5}(b), the increased analytics provision time is observed when a high value of $\beta$ is employed. As mentioned in H-NDAF, to improve the efficiency of the model storage, the subscribed models are stored at the model storage in advance than the requested models. Therefore, the request-based analytics cannot fully utilize the benefits of the model storage at the leaf NWDAF and they show prolonged provision time compared to subscription-based ones. By its definition, as $\beta$ increases, more requested-based analytics are considered and therefore the overall analytics provision time increases with the increase of $\beta$. This trend is apparent especially when the storage capacity is limited (e.g., $100$MB in Figure~\ref{fig:fig_5}(b)). If the storage capacity is affordable (e.g., $400$MB in Figure~\ref{fig:fig_5}(b)), the impact of $\beta$ is not significant. Consequently, it can be concluded that it is advantageous to use a subscription method to receive the analytics results in a short time.

\section{Open Research Issues}
\label{Sec:issues}

In this section, we present open research issues of H-NDAF for realizing more efficient network automation in 6G networks.

\subsection{Model Store Management}

In H-NDAF, we define simple criteria for storing models in the leaf NWDAF. Specifically, the leaf NWDAF prioritizes the subscribed models over the requested ones. This is because the subscribed models can be more regularly (and frequently) utilized, which improves the efficiency of the storage space. In addition, depending on the types of NFs, different levels of the frequency for the subscribed and requested models can be expected. Therefore, a more sophisticated caching strategy should be devised for the leaf NWDAF to efficiently manage its model store by considering these features.

\subsection{Distributed Training} 

In H-NDAF, inference loads can be distributed to leaf NWDAFs whereas training is performed only at the single root NWDAF. Therefore, if multiple model training tasks need to be conducted, the root NWDAF can be overloaded. To address this problem, distributed training with the help of leaf NWDAFs (e.g., federated learning (FL)~\cite{Li20}) can be exploited. That is, the leaf NWDAFs train local models, and the trained ones are aggregated at the root NWDAF as a global model. Even though FL can reduce the training load at the root NWDAF, there are several issues (i.e., straggler and non-independently and identically distributed data) when using FL in H-NDAF, which should be further investigated.

\subsection{Privacy and Security Management}

Privacy issues can occur when the root NWDAF collects raw data (especially UEs' personal information) from NFs. Specifically, an attacker can intercept the data in the middle of the root NWDAF and NFs. In addition, when delivering the trained model to the leaf NWDAF, the attacker can intercept the model and extracts the training data characteristics from that model~\cite{Basin18}. Meanwhile, an attacker can degrade the performance (e.g., accuracy and execution time) of the model. For example, when delivering the trained model from the root NWDAF to the leaf NWDAF, the attacker can change the parameters of the model (i.e., model poisoning attack) to reduce the model accuracy. Consequently, appropriate protection methods to the data and models should be incorporated with H-NDAF.

\section{Conclusion}
\label{Sec:Conclusion}

This article proposed H-NDAF for network automation of 6G core networks. In H-NDAF, the root NWDAF can construct the model providing accurate analytics with the global view. In addition, the leaf NWDAFs can provide analytics to their corresponding NFs in a timely manner. To show the feasibility of H-NDAF, we implemented H-NDAF using free5GC. Evaluation results demonstrate that H-NDAF provides sufficiently high accuracy and fast analytics provision time compared to the conventional frameworks. In our future work, we will extend H-NDAF to handle the presented open issues.

\section*{Acknowledgement}
\label{Sec:Acknowledgement}
This work was supported in part by Institute of Information \& communications Technology Planning \& Evaluation (IITP) grant funded by the Korea government (MSIT) (No. 2021-0-00739, Development of Distributed/Cooperative AI based 5G+ Network Data Analytics Functions and Control Technology) and in part by National Research Foundation (NRF) of Korea Grant funded by the Korean Government (MSIP) (No. 2021R1A4A3022102).

\begin{IEEEbiography}
[{\includegraphics[width=1in,height=1.25in,clip,keepaspectratio]
{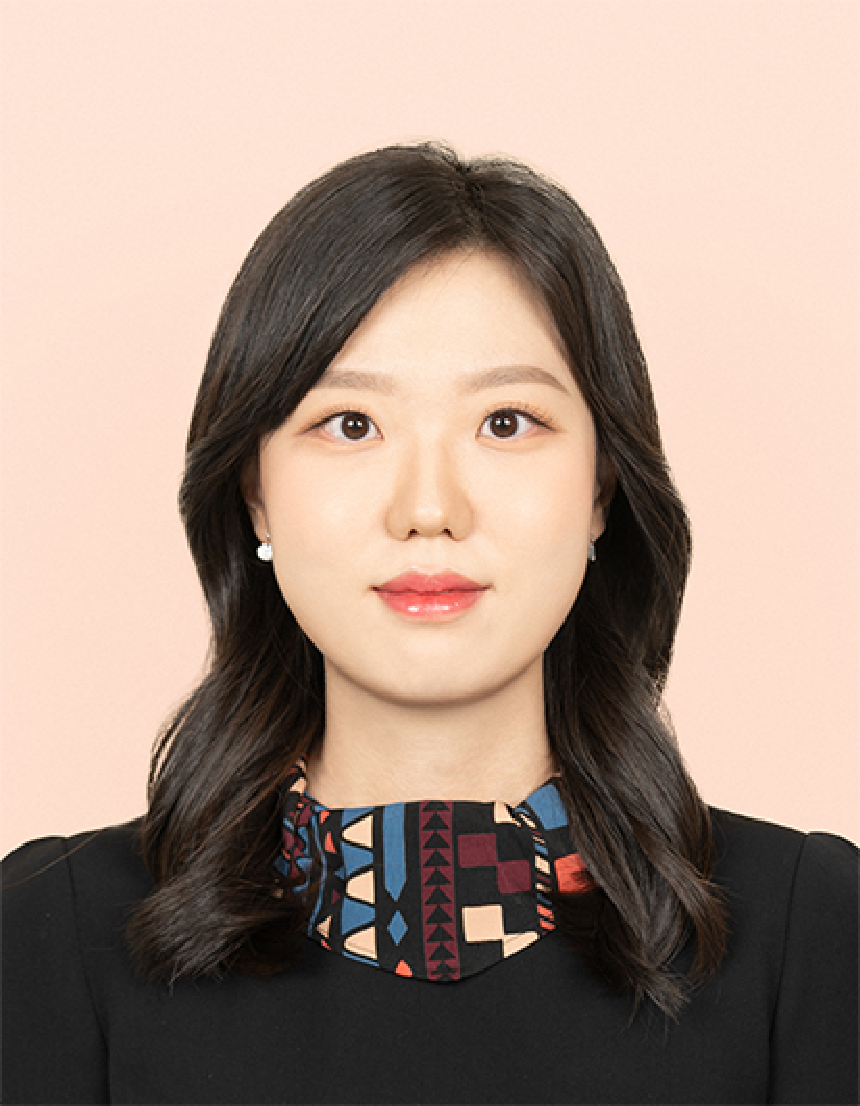}}]{YOUBIN JEON} (youbinee@korea.ac.kr) received the B.S. degree from Myongji University, Korea, in 2015. She is currently an M.S. and Ph.D. integrated course student in School of Electrical Engineering, Korea University, Korea. Her research interests include 5G/6G networks, network automation, robust model training, and ML/DL.
\end{IEEEbiography}

\begin{IEEEbiography}
[{\includegraphics[width=1in,height=1.25in,clip,keepaspectratio]
{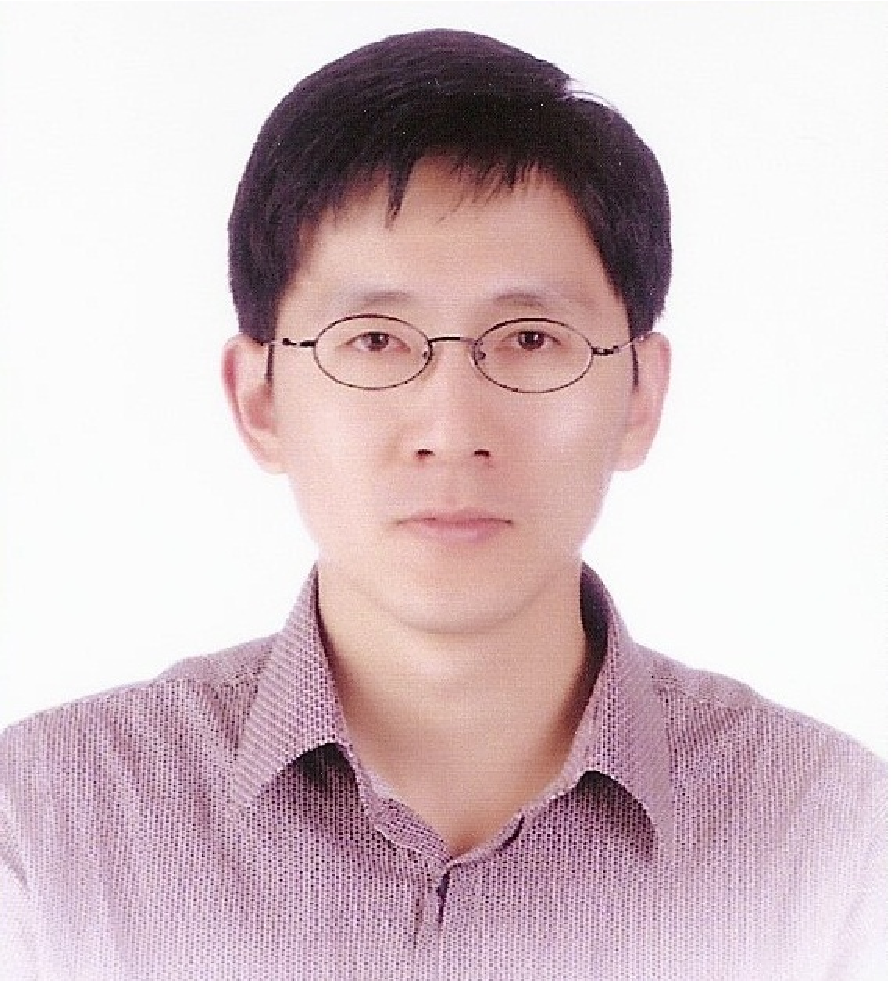}}]{SANGHEON PACK} [SM'11] (shpack@korea.ac.kr) received the B.S. and Ph.D. degrees from Seoul National University, Korea, in 2000 and 2005, respectively. In 2007, he joined the faculty of Korea University, Korea, He is currently a professor at the School of Electrical Engineering. His research interests include network softwarization and mobile edge computing.
\end{IEEEbiography}

\end{document}